Enrique Orduña-Malea and Rodrigo Costas

# 1. A scientometric-inspired framework to analyze EurekAlert! press releases

**Abstract:** Press releases about scholarly news are brief statements provided in advance to the press, including a description of the most relevant findings of one or more accepted scientific publications, usually under the condition that journalists will adhere to an embargo until the publication date. The existence of centralized platforms such as EurekAlert! allows press releases to be disseminated online as independent news articles. Press releases can include additional material (e.g., interviews, commentaries, explanatory tables, figures, media, recommended readings), which turn them into online objects with analytical value of their own. The objective of this work is to illustrate how press releases can be quantitatively analyzed applying similar tools and approaches as those applied in scientometric research (SCI). To achieve this goal, a scientometric-inspired analytical framework is proposed based on the formulation of spaces of interaction of objects, actors, and impacts. As such, the framework proposed considers press releases as science communication (SCO) objects, produced by different SCO actors (e.g., journalists), and the subject of receiving impact (e.g., tweets, links). To carry out this analysis, all press releases published by EurekAlert! from 1996 until 2021 (455,703 press releases), all tweets including at least one URL referring to a EurekAlert! press release (1,364,563 tweets), and all webpages with at least one URL referring to a EurekAlert! press release (54,089,233 webpages) have been studied. We argue that the large volume of press releases published and their online dissemination make these objects relevant in the measurement of SCO-SCI interactions.

## 1  Introduction

The communication of scientific results is an intrinsic part of the scientific endeavor. Scientists typically follow a formal communication channel regulated by a peer-review process. The basic unit of scholarly communication (communication of science within science) is the writing of a scientific publication which is published and publicly disseminated after peer review (Latour & Woolgar, 1986 [1979]). Other communication channels (Garvey, 1979) are also used to disseminate scientific research results in a more flexible way, and often also to broader audiences (e.g., preprints, scientific events, science fairs, distribution lists, media appearances, blogposts, museum exhibitions, etc.). The audience (scientists, mediators or the general public) can re-disseminate these research results under different forms to accomplish a number of goals, thus expanding the scope of what is commonly referred to as science communication (communication of science outside science).

Science communication is defined as the use of appropriate skills, media, activities, and dialog to enhance public awareness (e.g., creating familiarity with new aspects of science), enjoyment or other affective responses (e.g., appreciating science as entertainment or art), interest (e.g., voluntary involvement with science or its communication), opinion-forming (e.g., the forming, reforming, or confirming of science-related attitudes), and understanding (e.g., content, processes, and social factors) of science (Burns et al., 2003).

The increasing diversity of voices communicating science (Vogler & Schäfer, 2020) makes the role of science communication mediators (especially journalists) of special importance. News media routinely report on findings and discoveries newly published in scientific journals via science news (Kiernan, 2003; Groves et al., 2016).

News media facilitate the transmission of discoveries to citizens, enhancing thereby the public understanding of science (Autzen, 2014; Stockton, 2016) and favoring accountability of the public investment in science. Journalists can also act as watchdogs towards the influence of government and industry on scientists (de Vrieze, 2018). However, mass media are addressed to large unspecific audiences and a medialization of science effect can distort the scientific message (Weingart, 1998, 2012; Franzen, 2012; Franzen et al., 2012). The establishment of editorial decisions primarily based on news values and trendiness instead of the intrinsic publications' scientific value might favor the writing of new articles oversimplifying research results, exaggerating findings, hiding problematic information, or monitoring just a few highly selective multidisciplinary journals (Woloshin & Schwartz, 2002; Fanelli, 2012).

Media exposure could also alter the citation-based impact of publications (see also Chapters 4 and 5 within this volume). In this regard, the scientific literature has addressed two main hypotheses, the "publicity hypothesis," which assumes that media coverage genuinely increases the academic citations of the publications portrayed, and the "earmark hypothesis," which assumes that media coverage merely earmarks outstanding articles which would have received many citations anyway (Phillips et al., 1991). Earlier literature has proved that publications disseminated in the media obtained on average higher numbers of citations than those publications not covered by the media (Phillips et al., 1991; Kiernan, 2003; Fanelli, 2012; Alonso-Flores et al., 2020; Dumas-Mallet et al., 2020). Lemke et al. (2022) have also shown that articles mentioned in embargo e-mails receive higher citations. However, no causality has been established.

The scientific community in general – and science journal publishers in particular – have been traditionally reluctant to the informal dissemination of research results in the media, especially before peer review. The well-known Ingelfinger rule (the policy of considering a manuscript for publication only if it has not been submitted elsewhere, particularly through the popular press) constitutes a milestone in the control of science communication flows (Kiernan, 2006). Journal embargoes constitute another form of control on the dissemination of science news (Kiernan, 1997). To prevent the pre-publication of results in the media, journal editors alert science journalists about new original articles they deem remarkable a few days prior to the publication date (Franzen et al., 2012; Franzen, 2012). To do this, journals provide advance copies of the publications and distribute press releases (brief statements given to the press including a description of the most relevant findings of one or more accepted publications) under the condition that the journalists will adhere to a strict embargo until the publication date. This way, journal editors warrant that any new research has been properly peer-reviewed before being presented to the lay public, while they provide journalists enough time to write science news accurately (Stockton, 2016).

Web technologies allowed going one step further in the development of the embargo system, and online centralized platforms such as AlphaGalileo[1] or EurekAlert![2] were launched. These services allow not only journals but also other research bodies (mainly universities) to submit detailed press releases which will be delivered to the journalists subscribed to the service. These press releases are usually elaborated by trained journalists working in professional press offices and can include personal interviews with the authors and independent third-party commentaries, supplementary materials, recommended readings, and other informative elements. When the embargo for scientific publications has finished, and regardless of the re-dissemination of the embargoed information through media, the press releases are directly published online by the news service website and disseminated via social networking sites.

---

[1] https://www.alphagalileo.org/en-gb
[2] https://www.eurekalert.org

## 2 EurekAlert! as a press releases data source

EurekAlert! is a non-profit service established in May 1996, initially homed in the Stanford University servers and later moved and operated by the American Association for the Advancement of Science (AAAS) as a centralized online hub for science press releases (Stockton, 2016).

EurekAlert! only accepts contributions (press releases) from Public Information Officers (PIOs) at organizations that conduct, publish or fund scientific research in all scientific disciplines (there is no limit to the number of PIOs from one organization). These organizations must meet several eligibility criteria[3] and pay an annual fee. Journalists, who also need to meet specific eligibility criteria[4], might apply for free access to embargoed press releases submitted by PIOs. Only legitimate content owners may designate an embargo date/time for journalists when submitting a press release. EurekAlert! accepts specific press releases categories following specific restrictions. The scope and coverage of each press release category is provided in the supplementary material (Appendix A).

EurekAlert! includes specific channels including press releases written in French, German, Spanish, Portuguese, Japanese, and Chinese (which are also published in English), and holds specific news channels from science agencies (US Department of Energy, US National Institutes of Health, and US National Science Foundation) as well as general interest topics (climate change, cancer research or marine science). EurekAlert! has been questioned due to indirect effects both on journalists and PIOs (de Vrieze, 2018). First, some media have ended up replicating press releases without added value (Schäfer, 2017). Second, universities have professionalized their press offices and intensified their communication towards news media (Vogler & Schäfer, 2020), limiting the direct contact of scientists with journalists, and using press releases as effective means of communicating science and controlled tools to show utility (Autzen, 2014), overstating the societal implications of their findings (Franzen, 2012) to get attention and build reputation. These issues have raised concerns about the EurekAlert! model, which has come to be referred to as a marketplace (de Vrieze, 2018).

Scientometrics have paid little attention to EurekAlert! as an object of study. Bowman and Hassan (2019) developed the only descriptive analysis to date, describing the coverage of EurekAlert! in the Altmetric.com database. Lemke et al. (2021) analyzed qualitative aspects of EurekAlert! press releases (structure, linguistic accessibility, and the existence of narratives) and their potential influence on the impact of the publications promoted.

---

3 https://www.eurekalert.org/pioguidelines
4 https://www.eurekalert.org/reporterguidelines

While the authors acknowledge that science communication is not a single concept or construct, but a communication activity that involves multiple aspects and components that take place in multiple channels/modes from numerous sources, this study aims to examine press releases as specific quantifiable objects with the potential to enable the measurement of science communication interactions, an issue that has not been addressed so far.

## 3 A scientometric-inspired framework to study press releases

Press releases have been studied in the science communication (SCO) literature (Autzen, 2014), and recently as a source of news mentions (Ortega, 2020, 2021). However, there are no studies focused on the analysis of press releases as media objects with analytical value, and conceptually framed as spaces of interaction between media and science (Wouters et al., 2019; Costas et al., 2021). The main objective of this work is to fill this gap by illustrating how press releases (as particular SCO objects) can be quantitatively analyzed by applying similar tools and approaches as those applied in Scientometrics (SCI), using a scientometric-inspired framework based on the existence of spaces of interactions of objects, actors, and impacts.

In Table 1 we illustrate how the dimensions used to study SCI can to a large extent be related to SCO. In both cases there are objects that are being generated (e.g., journal articles, books or conference proceedings in SCI, and press releases, blogposts, or news items in SCO). There are also actors involved in the development of these objects (e.g., research authors or scientific journals in SCI, and science journalists or bloggers in SCO). Finally, there are also different impacts that can be quantitatively captured for both SCI and SCO. While scientometrics capture impacts between scientific articles (e.g., citation linkages among articles), there are also different types of impacts derived from SCO objects (e.g., press releases mentioning other press releases).

Finally, scientific actors and non-scientific actors can interplay both in SCI (e.g., mentions of tweets to scientific publications) and SCO (e.g., mentions of tweets to press releases). Precisely, based on the notion of "heterogeneous cou-pling" (Costas et al., 2021), it is theoretically possible to frame and capture interactions between SCI and SCO objects, actors and impacts. In the context of press releases, it can be argued that when a press release directly links to (or mentions) a scientific article, this represents a coupling between the two realms of SCO and SCI. The underlying idea is that these forms of heterogeneous couplings are of fundamental relevance, since they can unveil dynamics and interactions between these two realms, which could be accommodated in the scientometric-inspired framework proposed in Table 1.

**Table 1:** Scientometric-inspired framework to measure science communication.

| Dimension | Scientometrics (SCI) | | Science communication (SCO) | |
| --- | --- | --- | --- | --- |
| | Scope | Examples | Scope | Examples |
| Objects | Outputs produced in the scientific process | Articles; books; data; software; journals | Outputs produced in the science communication process, and outside the scientific process | Press releases; blogposts; streams; podcasts |
| Actors | Agents involved in the creation of SCI objects | Scientists; institutions; journals; academic publishers; funders | Agents involved in creation and dissemination of SCO objects | Science journalists; (social) media users; press offices; streamers; bloggers |
| Impacts | Impact of SCI objects on other SCI or SCO actors/objects | Citations between scientific articles; tweets mentioning papers | Impact of SCO objects on other SCO or SCI actors/objects | Tweets mentioning press releases; blogs mentioning press releases; press releases mentioning other press releases |

To discuss the potential of the scientometric-inspired framework described above, EurekAlert!, as the most comprehensive science press release distribution platform (Stockton, 2016), is taken as an illustrative case. First, a descriptive analysis is performed to present the main characteristics of press releases as SCO media objects. Furthermore, the online impact of press releases is studied through two online communication channels (the web at large and Twitter). Finally, the journals and publications mentioned in the EurekAlert! press releases (interactions) are also analyzed in order to determine which science is communicated by press releases, providing a basis for the further study of SCO-SCI heterogeneous couplings.

# 4 Methods

## 4.1 Press releases data

All press releases published by EurekAlert! since its inception were collected. Despite the existence of advance search features at the EurekAlert! website[5], the website does not support systematic large information retrieval options. As all press releases are published online as webpages under a specific URL address[6], it was decided to download all these documents directly via web crawling[7]. To do this, SocSciBot[8] v4 was configured to crawl all URLs under the "EurekAlert.org/press_release/" fold (ethical guidelines were followed by notifying the webmaster about the process, and by establishing one query per second to avoid crawling overload). This process was carried out during the first week of March 2021, obtaining a html copy of each webpage published until 28 February 2021. At the end of the process, 456,758 individual files were downloaded. A data cleansing process aimed at filtering out all those files not corresponding to a press release (e.g., sitemaps, forms, automatic server messages) was carried out, which yielded a final set of 455,703 press releases. The html file of each press release included all descriptive metadata fields that were created and curated by the EurekAlert! staff. A python script (see supplementary material) was subsequently written to extract the following metadata: keywords, description, date, funder, journal, type, institution, meeting, and region.

---

[5] https://www.eurekalert.org/search.php
[6] https://www.eurekalert.org/pub_releases/*
[7] While Eurekalert! provides a service for multilanguage press releases, covering Chinese, French, German, Japanese, Portuguese, and Spanish, the general URL of these press releases corresponds to the English version. This way, all online metrics gathered in the study are considering all language versions of each press release.
[8] http://socscibot.wlv.ac.uk

This python script was also used to extract all DOIs mentioned in each press release to characterize the publications covered by press releases, not included in the metadata fields. Several errors with DOIs were found (e.g., broken URLs, shortened URLs). These errors were systematically found for some journals (e.g., APL Photonics), which were corrected whenever possible. After this curation process, a total of 99,829 unique DOIs were found in 98,305 press releases (21.6% of all press releases), a percentage close to that found by Bowman and Hassan (2019), who also reported a small share (18%) of EurekAlert! press releases including DOIs.

Finally, data linking to the EurekAlert! press releases was gathered from Twitter (1,364,563 mentioning tweets) and the web at large via Majestic (54,089,233 linking webpages). A detailed description of the collection procedure is given below.

### 4.2 Twitter data

The Twitter API v2 was used to retrieve all tweets containing a URL leading to a press release published by EurekAlert! from 26 March 2006 (the day Twitter was launched) until 28 February 2021. To do this, the full-archive search endpoint, available through the Academic Research Product track, was used[9].

Despite some applications, such as Academic Mozdeh[10], already operating with this endpoint, these tools offer a predefined set of parameters out of all those available in the API. For these reasons, an ad hoc python script was written to query the Twitter full-archive search endpoint directly, using the following search query: url:"EurekAlert.org/press_release" -is:retweet

The full query (including dates of tweets publications, tweet-level metrics, and creator-level metrics) was debugged with Postman[11], and the script was launched in 15 April 2021 via a local server located at the university where the first author is affiliated. All data was obtained in a json file, which was subsequently distilled with OpenRefine[12] and finally exported to a spreadsheet to be

---

**9** https://developer.twitter.com/en/solutions/academic-research
**10** http://mozdeh.wlv.ac.uk/AcademicResearchTwitter.htm
**11** https://www.postman.com
**12** https://openrefine.org

statistically analyzed. This process gathered a total of 1,496,125 original tweets (retweets were excluded due to the computational complexity involved in their analysis) with at least one URL referring to a EurekAlert! press release. After a subsequent cleansing task, a total of 1,364,563 tweets including at least a URL pointing to each EurekAlert! press releases were obtained, which form the final Twitter dataset.

All links submitted within tweets are eventually wrapped with the t.co shortener[13]. These links are automatically un-shortened in the API response. However, when the link embedded in the tweet is already shortened (e.g. embedding a bit.ly URL in a tweet), the API response does not completely unshorten this URL (i.e., the API just unshortens from t.co to bit.ly). To solve this problem, another python script was developed to un-shorten all unresolved short URLs. After this process, a total of 260,780 unique press releases URLs were finally obtained.

## 4.3 Web data

Majestic Pro[14] was used to discover links referring to EurekAlert! press releases in the web at large. The historic index (which covers URLs crawled from 1 September 2015 to 1 April 2021 at the time of writing this manuscript) was used[15]. This link intelligence tool has been successfully used in the literature for webometrics analyses (Orduña-Malea, 2021).

Using "EurekAlert.org/press_releases/" as a seed URL path, the number of mentioning webpages (those webpages including at least one hyperlink to one specific URL under the URL seed path) and the number of mentioning websites (those websites including at least one hyperlink referring to one specific URL under the URL seed path) were gathered. As Majestic treats "http" and "https" URLs separately[16], a merging process was carried out to obtain all link-related metrics for each press release regardless of the protocol used.

Finally, Majestic's Flow Metrics (Citation Flow and Trust Flow) were obtained for each URL. These flow metrics are meant to capture some notion of the

---

[13] https://developer.twitter.com/en/docs/tco
[14] https://majestic.com
[15] Majestic offers two URL indexes, the Fresh index (URLs found during the last 120 days) and a historic index, a huge database covering the last five years.
[16] https://www.eurekalert.org/pub_releases/2015-12/aabu-paa122815.php and http://www.eurekalert.org/pub_releases/2015-12/aabu-paa122815.php are formally different URLs with independent metrics.

"prestige" or reputation of each URL through their linking webpages (Jones, 2012). Citation Flow is a score on a scale between 0 and 100 achieved by one website, based on the number of hyperlinks it receives. It measures how often a URL is linked. Therefore, it measures the quantity of links received. Trust Flow is a score on a scale between 0 and 100 achieved by one URL. It is based on the number of hyperlinks (and clicks on these links) from trusted seed sites that the URL receives. Therefore, it measures authority and ability to generate web traffic[17]. All data was extracted directly via the Majestic Pro interface as of 9 May 2021 and exported to a spreadsheet for statistical analysis. The overall process is summarized in Figure 1.

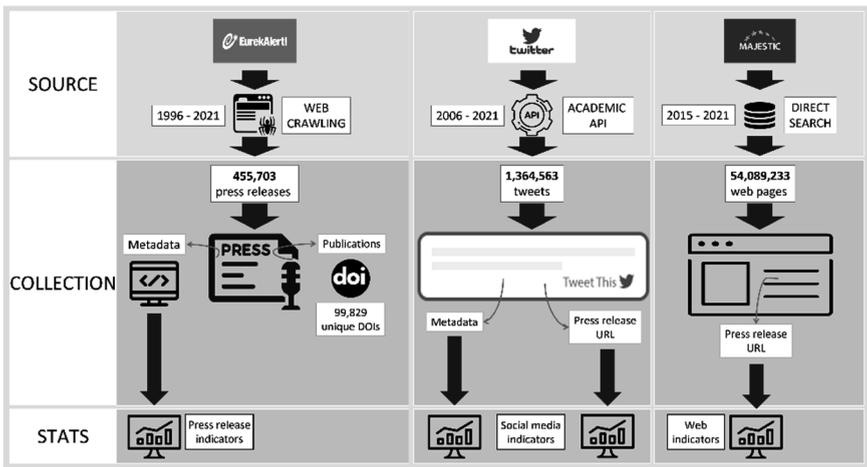

**Figure 1:** Diagram of the overall process (sources, data collection, and stats).

---

**17** The incorporation of these metrics only plays a role to illustrate the relevance of characterizing linking websites by their "prestige," but this does not represent a validation of this metric (which at best must happen in future research) nor a recommendation to be incorporated as a fixed element of the analytical framework proposed.

# 5 Results

## 5.1 The objects: EurekAlert! press releases

### 5.1.1 How many press releases are in EurekAlert!?

EurekAlert! has published 455,703 press releases online from its inception in 1996 until 28 February 2021[18]. This number has grown continuously, achieving a milestone of 35,232 press releases published in a single year in 2020 (Figure 2). This growth is evidenced by the fact that the publication output from 2016 to 2020 represents 30% of the total number of press releases published to date.

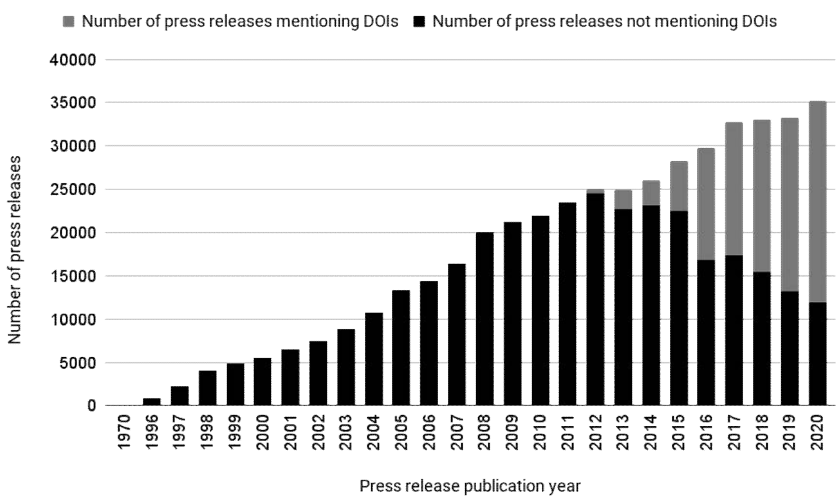

**Figure 2:** Annual number of press releases published in EurekAlert! (1996 to 2000).

### 5.1.2 Types of press releases in EurekAlert!

Most press releases are those of the research type (82.6%). The presence of business and grant announcement press releases (4.9% each) is also remarka-

---

**18** The crawling process discovered 10 press releases with an older publication date (both the URL and the documents publicly display 1969, while the html metadata of these documents display 1970). The authors associate these publication dates with human error.

ble, while the remaining categories are relatively infrequent (Table 2). The number of press releases published per day is not stable, the maximum having been detected on October 3 2016, when 228 press releases were published[19]. Autzen (2014) already detected a peak of press releases in 2013. Although this issue was referred to as exceptional, data shows the increasingly growth since then. The number of press releases mentioning a DOI is also displayed, evidencing an increasing growth. A total of 66.2% of all press releases published in 2020 included at least one DOI.

**Table 2:** Number of EurekAlert! press releases per publication type.

| Press Type | N | % |
| --- | --- | --- |
| Research | 376,199 | 82.6 |
| Business | 22,413 | 4.9 |
| Grant | 22,241 | 4.9 |
| Award | 14,743 | 3.2 |
| Meeting | 10,401 | 2.3 |
| Book | 4,433 | 1.0 |
| Media | 2,339 | 0.5 |
| Pubmeeting | 1,727 | 0.4 |
| Dissertation | 1,146 | 0.3 |
| Editorial | 60 | 0.0 |
| Total | 455,702 | 100.0 |

Note: 455,702 press releases with information in the "type" metadata field.

### 5.1.3 What is the thematic distribution of EurekAlert! press releases?

The predominance of medicine and health in the institutions submitting press releases is confirmed when analyzing the most frequently used terms included in the keywords field of each press release (Table 3). The term "medicine/health" appears in 41.2% of all press releases published.

---

[19] That day Yoshinori Ohsumi won the Nobel Prize for Medicine, but no other singular event has been identified.

Table 3: Keywords most frequently used by EurekAlert! press releases.

| Keyword | Number of occurrences |
| --- | --- |
| Medicine/health | 187,841 |
| Biology | 94,448 |
| Chemistry/physics/materials sciences | 48,227 |
| Cancer | 46,198 |
| Technology/engineering/computer science | 44,458 |
| Social/behavioral science | 39,569 |
| Public health | 39,061 |
| Genetics | 37,006 |
| Cell biology | 36,581 |
| Neurobiology | 33,423 |
| Earth science | 31,468 |
| Ecology/environment | 30,254 |
| Molecular biology | 24,890 |
| Cardiology | 23,582 |
| Behavior | 23,379 |
| Infectious/emerging diseases | 22,808 |
| Biochemistry | 22,172 |
| Biotechnology | 21,232 |
| Climate change | 20,388 |
| Mental health | 20,379 |

N = 455,564 press releases with information in the "description" metadata field.

The high predominance of medicine-related keywords may explain the limited vocabulary employed by EurekAlert! to describe press releases, as only 254 different keywords have been detected. The co-occurrence of all these keywords represents the scientific topics covered by EurekAlert! press releases over the years and their relations (Figure 3). A first cluster (green) represents medicine and health; a second cluster (red) represents social sciences; a third cluster (yellow) represents biology and ecology; a fourth cluster (blue) represents engineering; and a fifth cluster (purple) represents a mixture of physics, climate change, and oceanography. As expected, "medicine/health" is the node with the highest link strength (742,688). Other strongly connected terms are biology 457,648), computer science (226,258), and public health (225,447).

The collection of medicine-related items allows tailored thematic analysis using press releases as SCO object. Appendix B includes an illustrative case study of press releases mentioning "covid-19" or "coronavirus," showing how the application of the analytics can be expanded to include more specific and topical perspectives than the ones presented here.

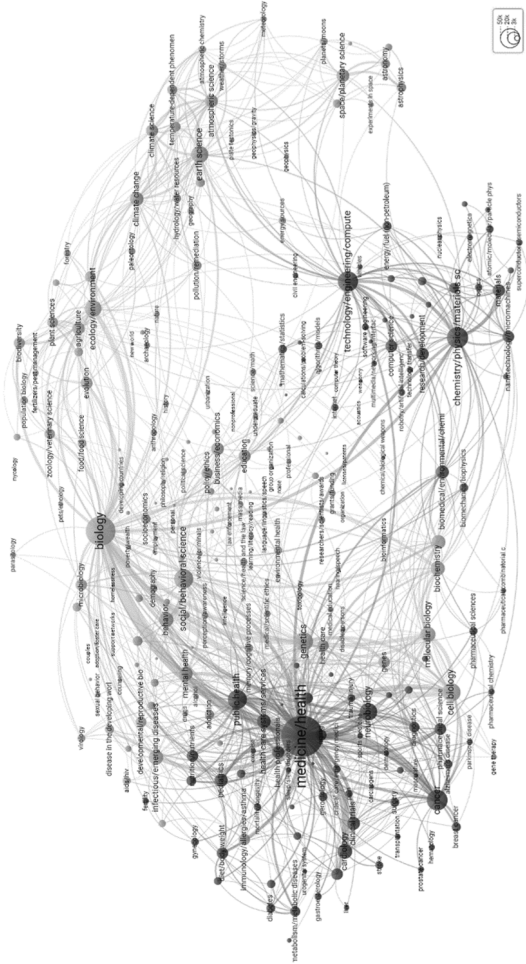

**Figure 3:** Map of co-occurrence of keywords in EurekAlert! press releases. Source: EurekAlert! data powered with VOSviewer (www.vosviewer.com). Map available online at: https://app.vosviewer.com/?json=https://drive.google.com/uc?id=1XhbI8m0lo8i5Ld6 SVrQXqD-dgK9wj1w5. N = 455,564 press releases with information in the "description" metadata field.

## 5.2 The actors: EurekAlert! press releases producers

### 5.2.1 Who are the most important producers of EurekAlert! press releases?

The Public Information Officers (PIOs) submitting press releases to the EurekAlert! are the most important actors in the generation of press releases. The origin of PIOs is strongly dominated by North American institutions (72.8% of all press releases submitted), followed by press releases submitted from European PIOs (21.8%). The presence of both African (0.7%) and especially South American (0.5%) institutions is marginal (Figure 4). There is a rather obvious distribution bias towards North American institutions, which can be explained by the fact that only institutions affiliated to EurekAlert! are eligible to submit press releases, and these mostly come from North American academic institutions.

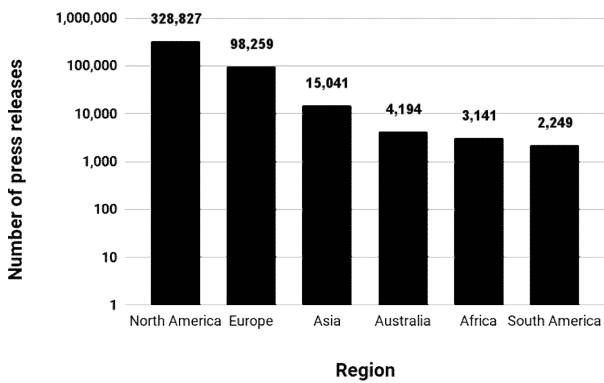

**Figure 4:** Number of press releases published in EurekAlert! per origin of PIOs. Note 1: 451,711 press releases with information in the "region" metadata field; note 2: y-axis in logarithmic scale.

The JAMA Network is found to be the institution with the greatest number of press releases submitted to EurekAlert! (7,333 press releases) followed by Goddard Space Flight Center–NASA space research laboratory in the United States (6,820), and the University of Texas (6,383). The results obtained are close to those obtained by Bowman and Hassan (2019). Supplementary material (Appendix C) includes the most active PIOs submitting press releases, organized into journals and publishers, associations and federations, universities and other institutions (e.g., hospitals not affiliated to universities, national institutes and government bodies). Despite associations acting as journal publishers, these entities have been categorized independently for the sake of clarity. Only in those cases when the institution marked as PIO by EurekAlert! corresponds to a specific entity's publication (e.g., Proceedings of the National Academy of Sciences) has it been included as a journal. In the case of universities, all internal units (e.g., medical centers, university presses, or research institutes) have been merged to obtain a unique value for universities as a type of institution.

These results show a remarkable presence of medical and health related journals and associations. Likewise, medical centers (e.g., University of Texas Health Science Center at Houston, Georgetown University Medical Center or Columbia University Medical Center) and schools (e.g., Johns Hopkins Medicine, Boston University School of Medicine or Michigan Medicine) constitute the most active EurekAlert! PIOs within universities.

## 5.3 The impact: mentions to EurekAlert! press releases

### 5.3.1 What is the online impact of EurekAlert! press releases?

Tweets and websites linking to EurekAlert! press releases represent forms of impact related to the press releases themselves (see Table 1). The online impact of press releases on social media has been measured via Twitter. A total of 1,364,563 tweets with at least a URL linking to a EurekAlert! press release have been analyzed. The number of mentioning tweets increases especially from 2010 onwards[20] and reaches a maximum by 2016 (238,881 tweets mentioning EurekAlert! press releases) (Figure 5).

Since 2016, the number of mentioning tweets has notably decreased, as well as the annual average of tweets per press release. A detailed analysis of the Twitter users linking to EurekAlert! press releases is available in the supplementary material (Appendix D). The online impact of press releases on the web at large has been measured via Majestic, through the 54,089,233 webpages with at least a URL linking to a EurekAlert! press release. The maximum value is observed in 2020, with 9,311,788 different mentioning webpages linking to press releases (Figure 6). The supplementary material (Appendix E) includes detailed information related to those domain names providing the most links to the EurekAlert! press releases.

---

**20** This effect is attributed to the creation of the EurekAlert! official Twitter account in September 2009.

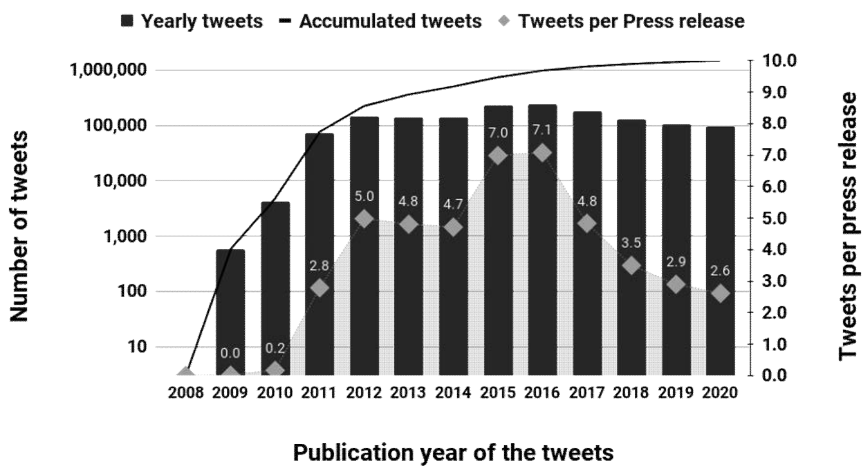

**Figure 5:** Number of tweets including a link to EurekAlert! press release per year. Note: tweets per press release considers tweets and press releases both published the same year. Note: y-axis (left) in logarithmic scale.

The analysis of all mentioning URLs extracted from Twitter and Majestic has revealed the existence of a sort of obsolescence effect, that is, the existence of active URLs not referring to press releases (i.e., outdated URLs). This effect is attributed to the fact that there are press releases that have been moved to a new URL or have been removed, while the old URL is still active, but referring to other contents[21]. This analysis has detected 907 outdated URLs in tweets and 2,039 outdated URLs in webpages.

The number of unique press releases being mentioned (either from Twitter or webpages) per press release publication year is offered in Table 4, excluding all outdated URLs. As we can observe, the coverage on Twitter is low until 2010-2011 (mainly attributed to the activity of the EurekAlert! official Twitter account [@EurekAlert], created in September 2009). Overall, 56.7% of all press releases have been mentioned at least once on Twitter[22]. From 2016-2017 onwards a decreasing trend is observed, which is in line with results previously displayed in Figure 5. The coverage of press releases on the web at large is otherwise significantly greater (79.1%).

---

[21] For example: https://www.eurekalert.org/pub_releases/2000-10/ASfM-Natd-2910100.php
[22] Twitter was launched in 2006. Therefore, it cannot be expected that many tweets will link to press releases published before this year.

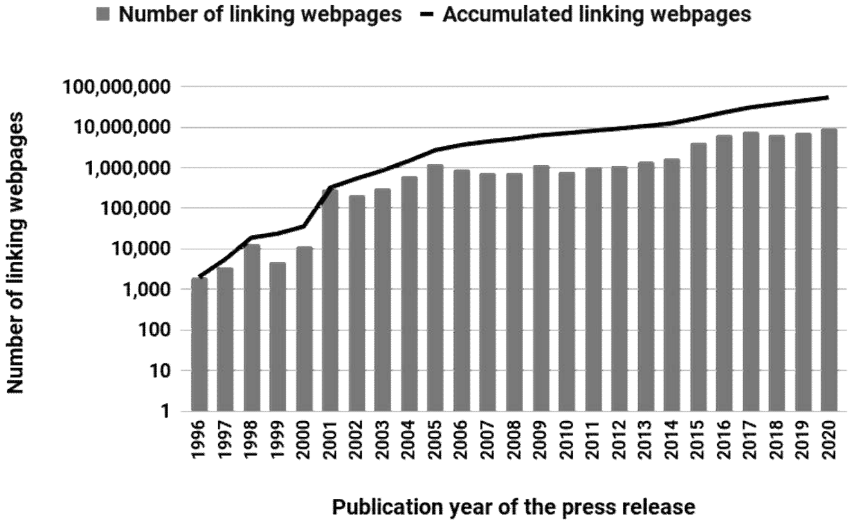

**Figure 6:** Number of webpages including a link to EurekAlert! press releases per year. Note: y-axis in logarithmic scale.

**Table 4:** Percentage of EurekAlert! press releases tweeted and web-linked (1996-2020).

| Press release publication year | Press releases published | Press releases tweeted | % | Press releases web-linked | % |
|---|---|---|---|---|---|
| 1996 | 798 | 5 | 0.63 | 115 | 14.4 |
| 1997 | 2,179 | 11 | 0.50 | 259 | 11.9 |
| 1998 | 4,072 | 19 | 0.47 | 506 | 12.4 |
| 1999 | 4,845 | 33 | 0.68 | 601 | 12.4 |
| 2000 | 5,536 | 27 | 0.49 | 763 | 13.8 |
| 2001 | 6,454 | 28 | 0.43 | 1,933 | 30.0 |
| 2002 | 7,491 | 68 | 0.91 | 2,966 | 39.6 |
| 2003 | 8,803 | 73 | 0.83 | 3,926 | 44.6 |
| 2004 | 10,780 | 94 | 0.87 | 5,512 | 51.1 |
| 2005 | 13,326 | 106 | 0.80 | 7,353 | 55.2 |
| 2006 | 14,362 | 183 | 1.27 | 8,964 | 62.4 |
| 2007 | 16,395 | 196 | 1.20 | 11,105 | 67.7 |
| 2008 | 20,037 | 296 | 1.48 | 14,245 | 71.1 |
| 2009 | 21,183 | 770 | 3.63 | 16,678 | 78.7 |
| 2010 | 21,977 | 3,644 | 16.58 | 17,777 | 80.9 |

| Press release publication year | Press releases published | Press releases tweeted | % | Press releases web-linked | % |
|---|---|---|---|---|---|
| 2011 | 23,466 | 13,325 | 56.78 | 21,357 | 91.0 |
| 2012 | 25,110 | 23,360 | 93.03 | 21,853 | 87.0 |
| 2013 | 24,961 | 23,235 | 93.09 | 23,424 | 93.8 |
| 2014 | 25,978 | 24,170 | 93.04 | 24,721 | 95.2 |
| 2015 | 28,258 | 26,766 | 94.72 | 27,665 | 97.9 |
| 2016 | 29,710 | 28,244 | 95.07 | 27,869 | 93.8 |
| 2017 | 32,732 | 29,859 | 91.22 | 30,445 | 93.0 |
| 2018 | 33,028 | 27,222 | 82.42 | 27,466 | 83.2 |
| 2019 | 33,232 | 26,210 | 78.87 | 29,539 | 88.9 |
| 2020 | 35,232 | 27,387 | 77.73 | 28,775 | 81.7 |
| TOTAL | 449,945 | 255,331 | | 355,817 | |

## 5.4 Heterogeneous couplings: press releases mentioning scientific articles

### 5.4.1 Which journals are covered in EurekAlert!?

The sources reported in EurekAlert! press releases can be seen as heterogeneous interactions between SCI objects (i.e., the journal) and SCO objects (i.e., the press release). A total of 12,071 unique scientific sources (including journals and conference proceedings) have been identified. The results obtained are also close to those obtained by Bowman and Hassan (2019), with the novelty of the increasing presence of the journal *Scientific Reports*. Table 5 includes the number of press releases where each publication source appears. In addition, we provide the number of publications with DOI published by each journal in the same period (1996 to 2020) according to Scopus. This way we can estimate what percentage of publications published by each scientific journal has been covered by EurekAlert! press releases.

Taking aside multidisciplinary journals (*PNAS, Science, and Nature*), a heavy representation of medicine and health journals (e.g., *British Medical Journal, Cell, JAMA, Lancet, New England Journal of Medicine* or *PLoS Medicine*) is found, which again confirms the propensity of covered research in this field on EurekAlert!. The significant percentage of publications from *PLoS Medicine* (47.9% of publications in the period covered) and Science Advances (43% of publications) is also noteworthy.[23]

Table 5: Number of EurekAlert! press releases per publication type.

| Journal | Number of publications with DOI | Number of press releases | % of publications covered in EurekAlert! |
| --- | ---: | ---: | ---: |
| PNAS | 90,160 | 15,840 | 17.6 |
| Science | 51,808 | 13,060 | 25.2 |
| Nature | 67,511 | 11,342 | 16.8 |
| Nature Communications | 35,503 | 7,818 | 22.0 |
| PLOS ONE | 241,450 | 6,777 | 2.8 |
| JAMA | 14,647 | 5,200 | 35.5 |
| Lancet | 48,483 | 4,416 | 9.1 |
| British Medical Journal of Medicine | 30,821 | 4,033 | 13.1 |
| Scientific Reports | 126,994 | 4,022 | 3.2 |
| New England Journal of Medicine | 32,687 | 3,497 | 10.7 |
| Cell | 12,673 | 3,140 | 24.8 |
| Journal of Clinical Investigation | 12,730 | 2,815 | 22.1 |
| Current Biology | 16,472 | 2,601 | 15.8 |
| Science Advances | 5,833 | 2,507 | 43.0 |
| Physical Review Letters | 81,488 | 2,400 | 2.9 |
| PLOS Medicine | 4,255 | 2,037 | 47.9 |
| Journal of Neuroscience | 31,188 | 2,033 | 6.5 |
| Neuron | 10,326 | 1,889 | 18.3 |
| Annals of Internal Medicine | 14,544 | 1,756 | 12.1 |
| Neurology | 28,701 | 1,686 | 5.9 |

N = 309,196 press releases with journal information.

---

[23] These percentages are even more significant if we consider that *PLoS Medicine* began operation in 2004 and *Scientific Advances* in 2015.

# 6 Discussion

This study illustrates how press releases can be quantitatively analyzed, applying similar tools and approaches as those applied in scientometric research. In this regard, the consideration of press releases as analytical objects with applicability to measure science communication interactions is discussed.

Press releases act as a filter of science by mentioning and promoting specific publications. As such we argue that press releases work as spaces of interaction between science and science communication. We assume that the selected publications might have specific value attributes, either academically oriented (i.e., new discoveries with great implications for research) and/or media-oriented (e.g., controversial results, topics of public interest). In either case, the study of these objects allows us to obtain a better understanding of science communication mechanisms due to the interactions established by press releases with other objects (scientific publications, journals, tweets, websites) and actors (PIOs, tweeters, and website authors). These types of interactions have not been measured so far. As the number of published press releases is growing (more than 30,000 items published annually by EurekAlert! alone), there arguably is a critical mass that permits large scale analyses, and which enables the introduction of advanced quantitative approaches to study SCI-SCO dynamics.

As the creation and dissemination processes of press releases follow different dynamics and purposes than scientific publications, the consideration of press releases as independent media objects with analytical value raises a series of peculiarities to consider. Below we discuss the most important data limitations of EurekAlert! as well as the main types of interactions captured through the proposed framework. We acknowledge that this is a first proposal of how to study press releases as science communication objects. Future research should focus on further developing this framework and its analytical scope.

## 6.1 EurekAlert! data limitations

Press releases metadata fields have remained the same since the launch of EurekAlert!, allowing the realization of quantitative analyses for the entire period of existence of EurekAlert!. However, most of the metadata fields are not harmonized. Consequently, institution and journal names appear under different variants, or typographic errors, which limit the quantitative analysis of the data without a substantial investment in data curation. The thematic keywords describing the scientific topics of press releases is one of the few controlled data elements in EurekAlert!. However, the vocabulary consists of only 254 keywords, which makes this controlled set of keywords somehow limited, particularly when describing new topics (e.g. COVID-19) or smaller fields. The expansion of the classification scheme used to describe press releases (e.g. including article-level data or expanding the keywords used) would provide EurekAlert! with a more dynamic and valuable tool to identify (and study) the topics of the press releases.

## 6.2 Objects, actors, and interactions

This study frames the investigation of press releases as spaces of interaction between science (SCI) and science communication (SCO). This could be seen as a derivation of the "heterogeneous couplings" framework proposed by Costas et al. (2020). As such, the framework proposed in this study considers press releases as SCO objects, produced by different SCO actors (e.g., PIOs, journalists), and receiving impact (e.g., tweets, links). Each SCO element (objects, actors, impacts) can in turn interact with other SCO and SCI elements. In this study we have illustrated how some of these objects, actors, and impacts related to SCO and SCI can be quantitatively captured and combined.

## 6.3 Objects: press releases as spaces of SCI-SCO interaction

Press releases can be seen as objects in the science communication process. At the same time, quite often, press releases interact with science objects (e.g., by linking directly to scientific publications). In some cases, these publications are linked as supplementary readings, but in other cases the press release is actively promoting these publications. This clearly illustrates the role of press releases as spaces of interaction between science communication (SCO) and science (SCI). In this study, we have illustrated firstly, how scientific publications are mentioned in press releases and secondly, how their own features (e.g., their journals of publication) can be further studied.

Only 21.6% of all EurekAlert! press releases include a link to at least one DOI. This low percentage is mainly attributed to the fact that EurekAlert! was launched long before the introduction of DOIs as the main standard to identify scientific publications. In those early years, publications were mentioned by URLs to journal websites without a DOI, or just via textual mention. The situation has changed over time, with more than 66% of EurekAlert! press releases mentioning a DOI in 2020. The increasing mention of DOIs in press releases opens up the possibility for more ambitious studies of SCO-SCI interactions, wherein features of press releases (e.g., content, producers, or received impact) can be studied in the context of the features of the scientific publications mentioned in the press release (e.g., scientific authors, scientific journal, covered topics, etc.).

## 6.4 Actors: press releases producers

Public Information Officers (PIOs) registered in EurekAlert! are typically the press release producers. This field may correspond to the journal or publisher where the article was published or the affiliation of one author of the scientific publication covered in that press release (generally but not necessarily the corresponding author). By measuring these actors, we can capture the activity of communication offices in promoting research. The specific publishers' active policies related to press releases submissions (and their investment in the operating of press offices) might introduce an inherent bias that should be considered.

The fact that large institutions may have different press offices operating makes institution-level analyses difficult. The existence of press offices for schools, departments or research centers may hinder the presence of the universities, while the existence of press offices for large editorials (e.g., Wiley or Elsevier) may hinder the presence of specific journals as PIOs.

EurekAlert! press releases show a strong North American bias in the coverage of science news. Like in the field of scientometrics when discussing the coverage and biases of its data sources (Martín-Martín et al., 2021; Visser et al., 2021), the coverage of press release sources (e.g., in the future also including AlphaGalileo and other science news platforms like The Conversation [Dudek & Costas, 2020]) can be seen as an additional important step in the development of quantitative studies of science communication.

## 6.5 Impact: mentions to press releases

Like scientific publications, which in scientometric approaches can be measured in terms of their impact either within science (e.g., citations among scientific publications) or outside of science (e.g., altmetrics), press releases may also generate further impact of their own. The existence of a distinctive URL for each press release allows this object to be used to measure impact-related events from which to generate a set of online metrics at the press release level. In this case, the mention of URLs both in websites and tweets has been studied.

With regard to Twitter, a very large set of tweets with at least one URL referring to a press release has been collected (1,364,563 tweets). However, only around half (56.7%) of all press releases have been tweeted at least once, and the percentage of annual tweets per press release has been decreasing since 2016, which might indicate a decline of Twitter as a communication channel for EurekAlert! press releases. In the case of the web at large, the coverage is much broader (79.1% of press releases have been linked at least once), receiving links from 54,089,233 webpages, mainly from organizations (.org top-level domain names), academic-related websites, and US universities.

Other forms of online impact (e.g., downloads, views, republication in other media) could also be seen as additional forms of the impact of press releases to be further investigated.

# 7 Conclusions

This study introduces a scientometric-inspired framework for the quantitative study of press releases as a novel information source for the study of SCO objects and their interactions with SCI elements. The large volume of press releases published and their wide online dissemination make these objects relevant in the measurement of SCO-SCI interactions.

Future research is recommended regarding the expansion of data sources related to SCO objects by considering other national and international press releases platforms, as well as monitoring other online channels of SCO dissemination (e.g., The Conversation). The ultimate ambition of the development of quantitative studies of press releases (and other SCO objects) is to gain a better understanding of SCI and SCO interactions. This understanding will enable the exploration of more ambitious questions regarding the communication of science and its effects on measuring the societal impact of SCI objects within the context of new emerging big data quantitative approaches, many of which have historically already been used to study science dynamics, such as scientometrics, webometrics and altmetrics.

# 8 Supplementary material

A dataset including scripts, raw data, and supplementary material is available at the following URL: https://riunet.upv.es/handle/10251/186769